\documentstyle[preprint,aps]{revtex}
\newcommand{\mathrm}[1]{\rm #1}
\tightenlines
\begin{document}

\draft
\title{The Nuclear Level Density and the Determination of Thermonuclear
Rates for Astrophysics}

\author{Thomas Rauscher}
\address{
Institut f\"ur Physik, Universit\"at Basel, Basel, 
Switzerland}

\author{Friedrich-Karl Thielemann}
\address{
Institut f\"ur Physik, Universit\"at Basel, Basel, Switzerland}

\author{Karl-Ludwig Kratz}
\address{
Institut f\"ur Kernchemie, Universit\"at Mainz,
Germany}
\maketitle

\begin{abstract}
\noindent
The prediction of cross sections for nuclei far off stability is crucial
in the field of nuclear astrophysics. In recent calculations the nuclear
level density -- as an important ingredient to the statistical model
(Hauser-Feshbach) --
has shown the highest uncertainties.
We present a global parametrization of nuclear level densities within the
back-shifted Fermi-gas formalism. Employment of an energy-dependent level
density parameter $a$, based on microscopic corrections from a recent FRDM mass
formula, and a backshift $\delta$, based on pairing and shell corrections,
leads to a highly improved fit of level densities at the
neutron-separation energy in the mass range $20\le A \le 245$. The
importance of using proper microscopic corrections from mass formulae is
emphasized. The resulting level description is well suited for
astrophysical applications.

The level density can also provide clues to the applicability of the
statistical model which is only correct for a high density of excited states.
Using the above description, one can derive a ``map''
for the applicability of the model to reactions of stable and unstable
nuclei with neutral and charged particles.
\end{abstract}


\section
{Introduction}

Explosive nuclear burning in astrophysical environments produces unstable 
nuclei, which again can be targets for subsequent reactions. In addition,
it involves a
very large number of stable nuclei, which are not fully explored
by experiments. Thus, it is  necessary to be able to predict reaction cross 
sections and thermonuclear rates with the aid of theoretical models.
Explosive burning in supernovae involves in general intermediate mass
and heavy nuclei. Due to a large nucleon number they have intrinsically
a high density of excited states. A high level density in the 
compound nucleus at the appropriate excitation energy allows to
make use of the statistical model approach for compound nuclear 
reactions (e.g.~\cite{HF52,MW79,GH92}), 
which averages over resonances. In this paper, we want to present new results
obtained within this approach and outline in a clear way, where its 
application is valid.  

It is often colloquially termed that the statistical model is 
only applicable for intermediate and heavy nuclei. However, the only necessary 
condition for its application is a large number of resonances
at the appropriate bombarding energies, so that the cross section can be
described by an average over resonances. This can in specific cases be valid for
light nuclei and on the other hand not be valid for intermediate mass nuclei
near magic numbers. Thus, another motivation of this investigation is to
explore the nuclear chart for reactions with a sufficiently
high level density, implying automatically that the
nucleus can equilibrate in the classical compound nucleus picture.

As the capture of an alpha particle leads usually to larger Q-values than
neutron or proton captures, the compound nucleus is created at a higher
excitation energy. Therefore, it
is often even possible to apply the Hauser-Feshbach formalism for light nuclei
in the case of alpha-captures.
Another advantage of alpha-captures is that the capture Q-values vary very 
little with the N/Z-ratio of a nucleus, for nuclei with Z$\le$50. For Z$>$50,
entering the regime of natural alpha-decay, very small alpha-capture Q-values
can be encountered for proton-rich nuclei. Such nuclei on the other hand do
not play a significant role in astrophysical environments, maybe with 
exception of the p-process.
This means that in the case of alpha-captures the requirement of large level 
densities at the bombarding energy is equally well fulfilled
at stability as for unstable nuclei. 
Opposite to the behavior for alpha-induced reactions,
the reaction Q-values for proton or neutron captures vary strongly
with the N/Z-ratio, leading eventually to vanishing Q-values at
the proton or neutron drip line. For small Q-values the compound
nucleus is created at low excitation energies and also for intermediate
nuclei the level density can be quite small. Therefore, it is not
advisable to apply the statistical model approach close to the 
proton or neutron drip lines for intermediate nuclei.
For neutron captures close to the neutron drip line in r-process
applications it might be still permissable for heavy and often
deformed nuclei, which have a high level density already at
very low excitation energies. 

In astrophysical applications usually different aspects are emphasized than 
in pure nuclear physics investigations. Many of
the latter in this long and well established field were focused on specific
reactions, where all or most "ingredients", like optical potentials for
particle and alpha transmission coefficients, level densities, resonance
energies and widths of giant resonances to be implementated in predicting
E1 and M1 gamma-transitions, were deduced from experiments. This of course,
as long as the statistical model prerequisites are met, will produce highly
accurate cross sections.

For the majority of nuclei in astrophysical applications such information
is not available. The real challenge is thus not the well established
statistical model, but rather to provide all these necessary ingredients
in as reliable a way as possible, also for nuclei where none of such 
informations are available. In addition, these approaches should be on a
similar level as e.g. mass models, where the investigation of hundreds
or thousands of nuclei is possible with managable computational effort,
which is not always the case for fully microscopic calculations.

The statistical model approach has been employed
in calculations of thermonuclear reaction rates for astrophysical 
purposes by many researchers~\cite{T66,M70,MF70,T72},
who in the beginning only
made use of ground state properties. Later, the
importance of excited states of the target was pointed out~\cite{A73}. 
The compilations~\cite{holm76,W78,thi87}
are presently the ones utilized
in large scale applications in all subfields of nuclear astrophysics,
when experimental information is unavailable.
Existing global optical potentials, mass models to predict Q-values,
deformations etc., but also the ingredients to describe giant resonance
properties have been quite successful in the past (see e.g. the review
by~\cite{cow91}). The major remaining uncertainty
in all existing calculations stems from the prediction of nuclear level
densities, which in earlier calculations gave uncertainties even beyond
a factor of 10 at the neutron separation energy~\cite{gil65},
about a factor of 8~\cite{W78}, and a factor of 5 even in
the most recent calculations (e.g.~\cite{thi87}; see 
Fig.3.16 in~\cite{cow91}). In nuclear reactions the 
transitions to lower lying states dominate due to the strong energy
dependence. Because the deviations are 
usually not as high yet at low excitation energies, the typical cross 
section uncertainties amounted to a smaller factor of 2--3.

We want to show in this paper, after a short description of the model and the 
required nuclear input, the implementation of a novel treatment of
level density descriptions~\cite{ilji92,igna78}, where
the level density parameter is energy dependent and shell effects vanish at
high excitation energies. 
This is still a phenomenological approach, making use of a back-shifted
Fermi-gas model, rather than a combinatorial approach based on microscopic 
single-particle levels. But it is the first one leading to a
reduction of the average cross section uncertainty
to a factor of about 1.4, i.e. an average deviation of about 40\% 
from experiments,
when only employing global predictions for all input parameters
and no specific experimental knowledge.
The degree of precision of the present approach will give astrophysical
nucleosynthesis calculations a much higher predictive power. In order to
give a guide for its application, we also provide a map of the nuclear chart 
which indicates where the statistical model requirements are fulfilled and 
its predictions therefore safe to use.

\section
{Thermonuclear Rates from Statistical Model Calculations} 

\subsection {The basic procedure}

A high level density in the compound nucleus permits to use averaged
transmission coefficients $T$, which do not reflect a resonance behavior,
but rather describe absorption via an imaginary part in the (optical)
nucleon-nucleus potential~\cite{MW79}.
This leads to the well known expression

\begin{eqnarray}
\sigma^{\mu \nu}_{i} (j,o;E_{ij})& = &
{{\pi \hbar^2 /(2 \mu_{ij} E_{ij})} \over {(2J^\mu_i+1)(2J_j+1)}} 
\nonumber \\
\label{cslab}
 & & \times \sum_{J,\pi} (2J+1){{T^\mu_j (E,J,\pi ,E^\mu_i,J^\mu_i,
\pi^\mu_i) T^\nu_o (E,J,\pi,E^\nu_m,J^\nu_m,\pi^\nu_m)} \over
{T_{tot} (E,J,\pi)}}
\end{eqnarray}
\par\noindent
for the reaction $i^\mu (j,o) m^\nu$ from the target
state $i^{\mu}$ to the exited state $m^{\nu}$ of the final nucleus, with a
center of mass energy E$_{ij}$ and reduced mass $\mu _{ij}$. $J$ denotes the
spin, $E$ the corresponding excitation energy in the compound nucleus 
and $\pi$ the parity of excited states.
When these properties are used  without subscripts they describe the compound
nucleus, subscripts refer to states of the participating nuclei in the
reaction $i^\mu (j,o) m^\nu$
and superscripts indicate the specific excited states. 
Experiments measure $\sum_{\nu} \sigma_{i} ^{0\nu} (j,o;E_{ij})$,
summed over all excited states of
the final nucleus, with the target in the ground state. Target states $\mu$ in
an astrophysical plasma are thermally populated and the astrophysical cross
section $\sigma^*_{i}(j,o)$ is given by
\begin{equation}
\label{csstar}
\sigma^*_{i} (j,o;E_{ij}) = {\sum_\mu (2J^\mu_i+1) \exp(-E^\mu_i /kT)
\sum_\nu \sigma^{\mu \nu}_{i}(j,o;E_{ij}) \over \sum_\mu (2J^\mu_i+1)
 \exp(-E^\mu_i/kT)}\quad.
\end{equation}
The summation over $\nu$ replaces $T_o^{\nu}(E,J,\pi)$ in Eq.~\ref{cslab} by
the total transmission coefficient
\begin{eqnarray}
T_o (E,J,\pi) & = &\sum^{\nu_m}_{\nu =0} 
T^\nu_o(E,J,\pi,E^\nu_m,J^\nu_m, \pi^\nu_m) \nonumber \\
\label{tot}
& &+ \int\limits_{E^{\nu_m}_m}^{E-S_{m,o}} \sum_{J_m,\pi_m}
T_o(E,J,\pi,E_m,J_m,\pi_m)\rho(E_m,J_m,\pi_m) dE_m\quad.
\end{eqnarray}
Here $S_{m,o}$ is the channel separation energy, and the summation over
excited 
states above the highest experimentally
known state $\nu_m$ is changed to an integration over the level density
$\rho$.
The summation over target states $\mu$ in Eq.~\ref{csstar} has to be generalized
accordingly. 

In addition to the ingredients required for Eq.~\ref{cslab}, like the
transmission coefficients for particles and photons, 
width fluctuation corrections $W(j,o,J,\pi)$ have to be
employed. They define the correlation factors with which all
partial channels for an incoming particle $j$ and outgoing particle $o$,
passing through the excited state $(E,J,\pi)$, have to be multiplied.
This takes into account that the decay of the state is not fully
statistical, but some memory of the way of formation is retained and
influences the available decay choices. The major effect is elastic
scattering, the incoming particle can be immediately re-emitted before
the nucleus equilibrates. Once the particle is absorbed and not
re-emitted in the very first (pre-compound) step, the equilibration is
very likely. This corresponds to enhancing the elastic channel by a
factor $W_j$. In order to conserve the total cross
section, the individual transmission coefficients in the outgoing
channels have to be renormalized to $T_j^\prime $. The total 
cross section is proportional to $T_j$ and, when summing over the elastic 
channel ($W_jT_j^\prime$) and all outgoing channels 
($T^\prime_{tot}-T^\prime_j$), one obtains the condition 
$T_j$=$T_j^\prime (W_jT_j^\prime/T^\prime_{tot})+T^\prime_j(T^\prime_{tot}
-T^\prime_j)/T^\prime_{tot}$. We can (almost) solve for $T^\prime_j$
\begin{equation}
\label{widthcorr}
T^\prime_j={T_j\over 1+ T^\prime_j(W_j-1)/T^\prime_{tot}}\quad.
\end{equation}
This requires an iterative solution for $T^\prime$ (starting in the
first iteration with $T_j$ and $T_{tot}$), which converges fast.
The enhancement factor $W_j$ has to be known in order to apply
Eq.~\ref{widthcorr}. A general expression 
in closed form was derived~\cite{V86}, but is computationally expensive to
use. A fit to results from Monte Carlo calculations gave~\cite{T74}
\begin{equation}
\label{newcorr}
W_j=1+{2\over 1+ T_j^{1/2}}\quad.
\end{equation}

For a general discussion of approximation methods see 
\cite{GH92,EP93}.
Eqs.~\ref{widthcorr} and \ref{newcorr} redefine the transmission
coefficients of Eq.~\ref{cslab} in such a manner that the total width is
redistributed by enhancing the elastic channel and weak channels over
the dominant one. Cross sections near threshold energies of new channel 
openings, where very different channel strengths exist, can only be
described correctly, when taking width fluctuation corrections into
account. Of the thermonuclear rates presently available
in the literature, only those by Thielemann et al.~\cite{thi87} include this
effect, but their level density treatment still contains large uncertainties. 
The width fluctuation
corrections of~\cite{T74} are only an approximation to the
correct treatment. However, it was shown that
they are  quite adequate~\cite{TZL86}.

The important ingredients of statistical model calculations as indicated in
Eqs.~\ref{cslab} through \ref{tot} 
are the particle and gamma-transmission coefficients $T$ and
the level density of excited states $\rho$. Therefore, the reliability of  
such calculations is determined by the accuracy with which these components 
can be evaluated (often for unstable nuclei). In the following we want to 
discuss
the methods utilized to estimate these quantities and recent improvements.

\subsection{Transmission Coefficients}
The transition from an excited state in the compound nucleus $(E,J,\pi)$
to the state $(E^\mu_i,J^\mu_i,\pi^\mu_i)$ in nucleus $i$ via the emission of
a particle $j$ is given by a summation over all quantum mechanically allowed
partial waves

$$T^\mu_j (E,J,\pi,E^\mu_i,J^\mu_i,\pi^\mu_i) =
\sum_{l=\vert J-s \vert}^{J+s}
\sum_{s=\vert J^\mu_i -J_j \vert}^{J^\mu_i + J_j}
T_{j_{ls}} (E^\mu_{ij}).  \eqno (6)$$
\par\noindent
Here the angular momentum $\vec l$ and the channel spin $\vec s =\vec J_j+
\vec J^\mu_i$ couple to $\vec J = \vec l +\vec s$. The transition energy
in channel $j$ is $E^\mu_{ij}$=$E-S_j-E^\mu_i$.

The individual particle transmission
coefficients $T_l$ are calculated by solving the Schr\"odinger equation
with an optical potential for the particle-nucleus interaction. All early
studies of thermonuclear reaction rates~\cite{T66,MF70,A73,T72,holm76,W78}
employed optical square well
potentials and made use of the black nucleus approximation. We employ 
the optical potential for neutrons and
protons given by~\cite{JLM77}, based on microscopic
infinite nuclear matter calculations for a given density,
applied with a local density approximation. 
It includes corrections of the imaginary part~\cite{F81,M82}.
The resulting s-wave neutron strength
function $<\Gamma^o/D> \vert_{\rm 1eV}=(1/2\pi )T_{n(l=0)}(\rm 1eV)$
is shown and discussed in~\cite{thi83a,cow91},
where several phenomenological optical potentials of the 
Woods-Saxon type and the equivalent square well potential used in earlier 
astrophysical applications are compared.
The purely theoretical approach gives the best fit. It is also expected to 
have the most reliable extrapolation properties for unstable nuclei.
A good overview on different approaches can be found in~\cite{V91}.

Deformed nuclei were treated in a very simplified way
by using an effective spherical potential of equal
volume, based on averaging the deformed potential over all possible
angles between the incoming particle and the orientation of the deformed
nucleus.

In most earlier compilations alpha particles were also treated 
by square well optical potentials. We employ a phenomenological Woods-Saxon 
potential~\cite{Mann78} based on extensive data~\cite{Mc66}.
For future use, for alpha particles and heavier projectiles,
it is clear that
the best results can probably be obtained with folding potentials (e.g.
\cite{C85,SL79,ohu}).

The gamma-transmission coefficients are treated as follows.
The dominant gamma-transitions (E1 and M1) have to be
included in the calculation of the total photon width.
The smaller, and therefore less important,
M1 transitions have usually been treated with the simple single particle
approach ($T \propto E^3$~\cite{BW52}), as also discussed in
\cite{holm76}. The E1 transitions
are usually calculated on the basis of the Lorentzian representation of the
Giant Dipole Resonance (GDR). Within this model, the E1 transmission
coefficient
for the transition emitting a photon of energy $E_{\gamma}$ in a nucleus
$^A_N Z$ is given by
\begin{equation}
\label{gamtrans}
T_{E1}(E_{\gamma}) = {8 \over 3} {NZ \over A} {e^2 \over \hbar c}
{ {1+\chi}
\over mc^2} \sum_{i=1}^2 {i \over 3} { {\Gamma_{G,i} E^4_\gamma} \over
{(E_\gamma^2 -E^2_{G,i})^2 + \Gamma^2_{G,i} E^2_\gamma}}\quad.
\end{equation}
Here $\chi(=0.2)$ accounts for the neutron-proton exchange 
contribution~\cite{LS89}
and the
summation over $i$ includes two terms which correspond to the split of the
GDR in statically deformed nuclei, with oscillations along (i=1) and
perpendicular (i=2) to the axis of rotational symmetry. Many microscopic
and macroscopic models have been devoted to the calculation of the GDR
energies ($E_{G}$) and widths ($\Gamma_G$).
Analytical fits as a function of A and Z were also used~\cite{holm76,W78}.
We make use of the (hydrodynamic) droplet model approach~\cite{My77} for $E_G$,
which gives an excellent fit to the GDR energies and
can also predict the split of the resonance for deformed nuclei, when
making use of the deformation, calculated within the droplet model.
In that case, the two resonance energies are related to the
mean value calculated by the relations~\cite{D58}
$E_{G,1}+2E_{G,2}=3E_G$, $E_{G,2}/E_{G,1}=0.911 \eta +0.089$.
$\eta$ is the ratio of the diameter along the nuclear
symmetry axis to the diameter perpendicular to it, and can be
obtained from the experimentally known deformation or mass model
predictions.

See~\cite{cow91} for a detailed description of the approach utilized
to calculate the gamma-transmission coefficients for the cross section
determination shown in this work.

\section{Level Densities}

\subsection{The Back-Shifted Fermi-Gas Model}
\label{fgas}
While the method as such is well seasoned,
considerable effort has been put into the improvement of the input for 
statistical Hauser-Feshbach models (e.g.\ \cite{cow91}). However, 
the nuclear level density has given rise to the largest uncertainties in 
the description of nuclear reactions~\cite{cow91,holm76,thi87,thi88}. For 
large scale astrophysical applications it is also necessary to not
 only find reliable 
methods for level density predictions, 
but also computationally feasible ones.

Such a model 
is the non-interacting Fermi-gas model~\cite{bethe36}. Most statistical 
model calculations use the back-shifted Fermi-gas 
description~\cite{gil65}. 
More sophisticated Monte Carlo shell model calculations~\cite{dean}, as well
as combinatorial approaches (see e.g.~\cite{paar}), have
shown excellent agreement with this phenomenological approach and justified
the application of the Fermi-gas description at and above the neutron
separation energy.
Here we want to apply
an energy-dependent level density parameter $a$ together with
microscopic corrections from nuclear mass models, which leads
to improved fits in the mass range $20\le A \le 245$.

Mostly the back-shifted Fermi-gas description, assuming an even 
distribution of odd and even parities (however, see e.g.~\cite{pich94} for 
doubts on the validity of this assumption at energies of astrophysical 
interest), is used~\cite{gil65}:

\begin{equation}
\rho(U,J,\pi)={1 \over 2} {\cal F}(U,J) \rho(U)\quad,
\end{equation}
with
\begin{eqnarray}
\rho(U)={1 \over \sqrt{2\pi} \sigma}{\sqrt{\pi} \over
12a^{1/4}}{\exp(2\sqrt{aU}) \over U^{5/4}}\ ,\qquad
{\cal F}(U,J)={2J+1 \over 2\sigma^2} \exp\left({-J(J+1) \over
2\sigma^2}\right) \\
\sigma^2={\Theta_{\mathrm{rigid}} \over \hbar^2} \sqrt{U \over a}\ ,\qquad
\Theta_{\mathrm{rigid}}={2 \over 5}m_{\mathrm{u}}AR^2\ ,\qquad
U=E-\delta\quad. \nonumber
\end{eqnarray}
The spin dependence ${\cal F}$ is determined by the spin cut-off parameter 
$\sigma$. Thus, the level density is dependent on only two parameters: 
the level density parameter $a$ and the backshift $\delta$, which 
determines the energy of the first excited state.

Within this framework, the quality of level density predictions depends 
on the reliability of systematic estimates of $a$ and $\delta$. The 
first compilation for a large number of nuclei was provided 
by~\cite{gil65}. They found that the backshift $\delta$ is well 
reproduced by experimental pairing corrections. They also were the first 
to identify an empirical correlation with experimental shell corrections 
$S(N,Z)$
\begin{equation}
\label{aovera}
{a \over A}=c_0+c_1S(N,Z)\quad,
\end{equation}
where $S(N,Z)$ is negative near closed shells. Since then, a number of 
compilations have been published and also slightly different functional 
dependencies have been proposed (for references, see e.g.~\cite{cow91}), 
but they did not necessarily lead to better predictive power.

Improved agreement with experimental data was found~\cite{thi87,thi88} by 
dividing the nuclei into three classes [(i) those within three units of 
magic nucleon numbers, (ii) other spherical nuclei, (iii) deformed 
nuclei] and fitting separate coefficients $c_0$, $c_1$ for each class. In that 
case the mass formula in Ref.~\cite{Hilf76}
was used. For the backshift $\delta$ the description
\begin{equation}
\delta=\Delta(Z,N)
\end{equation}
was employed, deriving $\Delta(Z,N)$ from the pairing correlation of a 
droplet model nuclear mass formula with the values
\begin{eqnarray}
\label{pairing}
\Delta_{\mathrm{even-even}}&=&{12 \over \sqrt{A}}\quad,\nonumber\\
\Delta_{\mathrm{odd}}&=&0\quad,\\
\Delta_{\mathrm{odd-odd}}&=&-{12 \over \sqrt{A}}\quad.\nonumber
\end{eqnarray}
With this treatment smaller deviations were found, compared to previous 
attempts~\cite{gil65,holm76}. However, the number of parameters was 
considerably increased at the same time.

The back-shifted Fermi-gas approach diverges for $U=0$ (i.e.\ $E=\delta$, if
$\delta$ is a positive backshift). In order to get the correct behavior at
very low excitation energies, the Fermi-gas description
can be combined with the constant temperature formula (\cite{gil65};
\cite{GH92} and references therein)
\begin{equation}
\label{ctemp}
\rho(U) \propto {\exp(U/T) \over T}\quad.
\end{equation}
The two formulations are matched by a tangential fit determining $T$.

\subsection{Thermal Damping of Shell Effects}
\noindent
An improved approach has to consider the energy dependence of the shell 
effects which are known to vanish at high excitation energies~\cite{ilji92}.
Although, for astrophysical purposes only energies close to the particle 
separation thresholds have to be considered, an energy dependence can 
lead to a considerable improvement of the global fit. This is especially 
true for strongly bound nuclei close to magic numbers.

An excitation-energy dependent description was initially proposed 
by~\cite{igna75,igna78}
for the level density parameter $a$:
\begin{equation}
\label{endepa}
a(U,Z,N)=\tilde{a}(A)\left[1+C(Z,N){f(U) \over 
U}\right]\quad,
\end{equation}
where
\begin{equation}
\tilde{a}(A)=\alpha A+\beta A^{2/3}
\end{equation}
and
\begin{equation}
f(U)=1-\exp(-\gamma U)\quad.
\end{equation}
The values of the free parameters $\alpha$, $\beta$ and $\gamma$ are
determined by fitting to experimental level density data.

The shape of the function $f(U)$ permits the two extremes: (i) for small
excitation energies the original form of Eq.~\ref{aovera} is retained
with $S(Z,N)$ being replaced by $C(Z,N)$, (ii)
for high excitation energies $a$/$A$ approaches the continuum value obtained
for infinite nuclear matter.
Previous attempts to find a global description of level densities used 
shell corrections $S$ derived from comparison of liquid-drop 
masses with experiment ($S\equiv M_{\mathrm{exp}}-M_{\mathrm{LD}}$) or 
the ``empirical'' shell corrections $S(Z,N)$ given by~\cite{gil65}.
A problem 
connected with the use of liquid-drop masses arises from the fact that 
there are different liquid-drop model parametrizations available in the
literature which produce quite different values for $S$~\cite{meng94}.

However, in addition the meaning of the correction parameter inserted into
the level density formula (Eq.~\ref{endepa}) has to be reconsidered. 
The fact that nuclei approach a spherical shape at high excitation energies
has to be included.
Actually, the correction parameter $C$ should 
describe properties of a nucleus differing from the {\it spherical} macroscopic 
energy and include terms which are vanishing at 
higher excitation energies. The latter requirement is mimicked by the form
of Eq.~\ref{endepa}.
Therefore, the parameter should rather be identified with 
the so-called ``microscopic'' correction $E_{\mathrm{mic}}$ than with the
shell correction. The mass of 
a nucleus with deformation $\epsilon$ can then be written as~\cite{moell95}
\begin{equation}
\label{emic}
M(\epsilon)=E_{\mathrm{mic}}(\epsilon)+E_{\mathrm{mac}}
(\mathrm{spherical})\quad.
\end{equation}
Alternatively, one can write 
\begin{equation}
M(\epsilon)=E_{\mathrm{mac}}(\epsilon)+E_{\mathrm{s+p}}
(\epsilon)\quad,
\end{equation}
with $E_{\mathrm{s+p}}$ being the shell-plus-pairing correction.
The confusion about the term ``microscopic correction'', which is 
sometimes used in an ambiguous way, is also pointed out 
in~\cite{moell95}. 
Thus, the above mentioned ambiguity 
follows from the inclusion of deformation-dependent effects
into the macroscopic part of the mass formula.

Another important ingredient is
the pairing gap $\Delta$, related to the backshift $\delta$. 
Instead of assuming constant pairing (cf.\ \cite{Reisdorf}) 
or a fixed dependence on the mass
number $A$ (cf.\ Eq.~\ref{pairing}), we 
determine the pairing gap $\Delta$ from differences in the binding
energies (or mass differences, respectively) of
neighboring nuclei.
Thus, for the neutron pairing gap $\Delta_{\mathrm{n}}$ one
obtains~\cite{rong92}
\begin{equation}
\label{pair}
\Delta_{\mathrm{n}}(Z,N)={1 \over 2} \left[ 
2E^G(Z,N)-E^G(Z,N-1)-E^G(Z,N+1)\right]\quad,
\end{equation}
where $E^G(Z,N)$ is the binding energy of the nucleus $(Z,N)$.
Similarly, the proton pairing gap $\Delta_{\mathrm{p}}$ can be 
calculated.

At low energies, this description is again combined with the constant
temperature formula (Eq.~\ref{ctemp}) as described above.

\subsection{Results}
\noindent
In our study we utilized 
the microscopic correction of
a most recent mass formula~\cite{moell95}, calculated with the Finite Range 
Droplet Model FRDM
(using a folded Yukawa shell model with Lipkin-Nogami pairing)
in order to determine the parameter $C(Z,N)$=$E_{\mathrm{mic}}$. The 
backshift $\delta$ was calculated by 
setting $\delta(Z,N)$=1/2$\{\Delta_{\mathrm{n}}(Z,N)+\Delta_{\mathrm{p}}(Z,N)\}$
and using Eq.~\ref{pair}.
In order to obtain the parameters $\alpha$, $\beta$, and $\gamma$, we 
performed a fit to 
experimental data on s-wave neutron resonance spacings of 272 nuclei at 
the neutron separation energy. The 
data were taken from a recent compilation~\cite{ilji92}.
Another recent investigation~\cite{meng94} also attempted to fit level density 
parameters, but made use of a slightly different description of the energy 
dependence of $a$ and different pairing gaps.

As a quantitative overall estimate of the agreement between calculation 
and experiment, one usually quotes the averaged 
ratio~\cite{gil65,ilji92}
\begin{equation}
g\equiv \left< {\rho_{\mathrm{calc}} \over \rho_{\mathrm{exp}}}\right> =
\exp \left[{1 \over n} \sum_{i=1}^{n}\left( \ln {\rho_{\mathrm{calc}}^i
\over \rho_{\mathrm{exp}}^i} \right)^2 \right]^{1/2}\quad,
\end{equation}
with $n$ being the number of nuclei for which level densities 
$\rho$ are experimentally known.

As best fit 
we obtain an averaged ratio $g=1.48$ with the parameter 
values $\alpha=0.1337$, $\beta=-0.06571$, $\gamma=0.04884$. 
The ratios of 
experimental to predicted level densities (i.e. theoretical to experimental
level spacings) for the nuclei considered are
shown in Fig.~\ref{figrat}. As can be seen, for the majority of nuclei 
the absolute deviation is less than a factor of 2. This is a 
satisfactory improvement over the theoretical level densities previously 
used in astrophysical cross section calculations, where deviations of a 
factor 3--4~\cite{thi87,thi88}, or even in excess of a factor of 
10~\cite{cow91,holm76} were found. Such a 
direct comparison was rarely shown in earlier work. Mostly the level
density parameter $a$, entering exponentially into the level density, was
displayed.
Closely examining the nuclei with the 
largest deviations in our fit, we were not able to find any remaining
correlation of the deviation with separation energy (i.e. excitation
energy) or spin.

Although we quoted the value of the parameter $\beta$ above (and will do so
below) as we left it as an open parameter in our fits, one can see that it
is always small and can be set to zero without considerable increase in the
obtained deviation. Therefore, it is obvious that actually only two
parameters are needed for the level density description.

As an alternative to the FRDM mass formula~\cite{moell95}, in 
Fig.~\ref{hilffit} we show the results
when making use of the well-known mass formula by Hilf et al.~\cite{Hilf76}
which turned out to be successful in predicting properties of nuclei
at and close to stability. The parameter set $\alpha=0.0987$, 
$\beta=0.09659$, $\gamma=0.05368$ yields an averaged ratio of $g=2.08$.
It can be seen from Fig.~\ref{hilffit} that not only the average
scatter is somewhat larger than with the FRDM input, but also that
this mass formula has problems in the higher mass regions. Only an
artificial alteration by about $-3$ MeV or more of the microscopic term in the
deformed mass regions $80 \le N \le 86$ and $103 \le N \le 113$ can
slightly improve the fit but the remaining scatter still leads to
$g=1.85$. The difference in the calculated level density from the FRDM
and the Hilf mass formulae is plotted in Fig.~\ref{hcomp}.
The latter mass formula leads to a significantly higher level density 
(by about a factor of 10) around 
the neutron magic number $N=82$, whereas the level density remains lower
(by a factor of 0.07) close to the drip lines for $N>115$.

A fit comparable to the quality of the FRDM approach can be obtained when
employing a mass formula from an extended Thomas-Fermi plus Strutinsky integral
model (ETFSI)~\cite{pearson}. In order to extract a microscopic correction for
this already microscopic calculation, we subtracted the FRDM
spherical macroscopic part $E_{\mathrm{mac}}$ (see Eq.~\ref{emic}) from
the ETFSI mass and took this difference to be the ETFSI microscopic correction.
The pairing gaps were calculated as described above. This leads to a fit
with $g=1.61$, yielding the parameter values $\alpha=0.12682$,
$\beta=-0.03652$ and $\gamma=0.045$. However, although the fit is closer to
the one obtained with FRDM than the Hilf result, the deviations for
unstable nuclei are somewhat larger. The maximum deviation is a factor of
about 38 for ETFSI, as compared to a factor of 16 for the Hilf approach.
Both formulae yield lower level densities than the FRDM for nuclei close to
the dripline with
$N>130$ and higher level densities for neutron rich nuclei close to the
magic shell at $N=82$. The ratios of the level density from the ETFSI
approach to those of the FRDM are shown in Fig.~\ref{etfsicomp}.

Different combinations of masses and microscopic corrections from other 
models (droplet model by Myers and Swiatecki~\cite{MyS}, 
Cameron-Elkin mass formula 
and shell corrections~\cite{Camel}) were also tried but did not 
lead to better results.
Our fit to experimental level densities is also better than a recent
analytical BCS approach~\cite{gor96a,gor96b} which tried to
implement level spacings from the ETFSI model in a consistent fashion.

To see the effect of the new level density description (utilizing FRDM input)
on the calculated
cross sections, 30 keV neutron capture cross sections from 
experiment~\cite{bakanew} are compared to our calculations in 
Fig.~\ref{bakaplot}.
Plotted are only nuclei for which the statistical model can be applied to
calculate the cross section, using the criteria derived in the next section.
An improvement in the overall deviation can be seen, compared to previous
calculations~\cite{cow91}. However, one systematic deviation can clearly
be seen in the $A\simeq90$ region. That ``peak'' is not caused by a
deficiency in the general level density description but by the microscopic
input. The FRDM model overestimates the microscopic corrections close to
the $N=50$ shell~\cite{moell95}.

We see that the uncertainty in level density translates into a similar
uncertainty of the neutron capture cross sections which are used here as a
representative example for applications to capture cross sections. Although
this does not seem to be a dramatic improvement for the experimental
cross sections of stable nuclei over the previous approach~\cite{thi87,thi88},
the purely empirical and also artificial division of nuclei into three
classes of level density treatments could be avoided. The reason is that
the excitation energy dependence was treated in the generalized ansatz
of~\cite{igna78}, ensuring the correct energy dependence which will also
yield correct results when the adjustment is not done at the typical
separation energy of 8--12 MeV for stable nuclei but also for nuclei far
from stability with smaller separation energies.

The remaining uncertainty in the extrapolation is the reliability far off 
stability of
the nuclear-structure model from which the microscopic corrections and
pairing gaps (and the masses) are taken.
However, recent investigations in 
astrophysics and nuclear physics have shown the robustness of the 
FRDM approach~\cite{irgendwas}. Recently improved purely 
microscopic models have exhibited similar behavior towards the drip 
lines~\cite{doba} but there are no large scale calculations over the 
whole chart of nuclei available yet which include deformation. Therefore, 
the FRDM model used here is among the most reliable ones available at present.

\section{Applicability of the Statistical Model}

\noindent
Having a reliable level density description also permits to analyze
when and where the statistical model approach is valid. 
Generally speaking, in order to apply the model correctly a sufficiently 
large number of levels in the compound nucleus is needed in the relevant 
energy range which can
act as doorway states to forming a compound nucleus.
In the 
following this is discussed for neutron-, proton- and alpha-induced reactions
with the aid of the level density approach presented above.
This section is intended to be a guide to a meaningful and correct 
application of the statistical model.

The nuclear reaction rate per particle pair at a given stellar temperature
$T$ is determined by folding the reaction cross section with the
Maxwell-Boltzmann (MB) velocity distribution of the projectiles~\cite{fowler}
\begin{equation}
\left< \sigma v \right>=\left( \frac{8}{\pi \mu} \right) ^{1/2}
\frac{1}{\left( kT \right) ^{3/2}}
\int_0^{\infty} \sigma(E) E \exp \left( -\frac{E}{kT} \right) dE \quad.
\end{equation}
Two cases have to be considered: Reactions between charged particles and
with neutrons.

\subsection{The Effective Energy Window}

\noindent
The nuclear cross section for charged particles is strongely suppressed
at low energies due to the Coulomb barrier.
For particles having energies less than the height of the Coulomb barrier,
the product of the penetration factor and the MB distribution function
at a given
temperature results in the
so-called Gamow peak, in which most of the reactions will
take place~\cite{rolfs}. Location and width of the Gamow peak depend 
on the charges
of projectile and target, and on the temperature of the interacting plasma.

When introducing the astrophysical $S$ factor $S(E)=\sigma(E)E\exp(2\pi
\eta)$ (with $\eta$ being the Sommerfeld parameter), one can easily see
the two contributions of the velocity distribution and the penetrability
in the integral:
\begin{equation}
< \sigma v >=\left( \frac{8}{\pi \mu} \right) ^{1/2}
\frac{1}{\left( kT \right) ^{3/2}}
\int_0^{\infty} S(E) \exp \left[ - \frac{E}{kT} - \frac{b}{E^{1/2}} 
\right] \quad,
\end{equation}
where the quantity $b=(2 \mu)^{1/2} \pi e^2 Z_1 Z_2 / \hbar$ arises from
the barrier penetrability. Taking the first derivative of the integrand
yields the location of the Gamov peak $E_0$~\cite{rolfs,fowler},
\begin{equation}
E_0= \left( \frac{bkT}{2} \right) ^{2/3}=1.22(Z_1^2 Z_2^2 \mu T_6^2)^{1/3}
\, \mathrm{keV}\quad,
\end{equation}
with the charges $Z_1$, $Z_2$ and the reduced mass $\mu$ of the involved
nuclei at a temperature $T_6$ given in 10$^6$ K. The effective width
$\Delta$ of the energy window can be derived as
\begin{equation}
\Delta=\frac{16E_0 kT}{3}^{1/2}=0.749 (Z_1^2 Z_2^2 \mu T_6^5)^{1/6}
\, \mathrm{keV}\quad.
\end{equation}


In the case of neutron-induced reactions 
the effective energy window has to be derived in a slightly different
way. For s-wave neutrons ($l=0$) the energy window is simply given by the
location and width of the peak of the MB distribution function.
For higher partial waves the penetrability of the centrifugal barrier
shifts the effective energy $E_0$ to higher energies, similar to the
Gamov peak. For neutrons with energies less than the height of the
centrifugal barrier this can be approximated by~\cite{wagoner}
\begin{eqnarray}
E_0\approx 0.172T_9\left( l+{1 \over 2}\right)\quad \mathrm{MeV,}\\
\Delta \approx 0.194T_9\left( l+{1 \over 2}\right)^{1/2}\quad \mathrm{MeV.}
\end{eqnarray}
\noindent
The energy $E_0$ will always be comparatively close to the neutron
separation energy.

\subsection{The Criterion for the Application of the Statistical Model}

\noindent
Using the above effective energy windows for charged and neutral
particle reactions, a criterion for the 
applicability can be derived from the level density. 
For a reliable application of the statistical model a sufficient number
of nuclear levels has to be within the energy window, thus contributing
to the reaction rate. For narrow, isolated resonances, the cross sections
(and also the reaction rates) can be represented by a sum over individual
Breit-Wigner terms. At higher energies, with increasing level density, the
sum over resonances may be approximated by an integral over $E$~\cite{wagoner}.

Numerical test calculations were made in order to find the average number of 
levels per energy window which is sufficient to allow the substitution of
the sum by an integral over the HF cross section. Fig.~\ref{levcrit} shows 
the dependence of the
ratio between sum and integral~\cite{wagoner} on the number of levels in
the energy window. To achieve 20\% accuracy, about 10 levels are needed in
the worst case (non-overlapping, narrow resonances). Usually, neutron s-wave
resonances are comparatively broad and thus a smaller number of levels
could be sufficient.
However, applying the statistical model (i.e. integrating over a level
density instead of summing up over levels) for a level density which is
not sufficiently large, results in
an overestimation of the actual cross section, as can be seen
in Fig.~\ref{levcrit} and was also shown in Ref.~\cite{laura}.
Therefore, in the following we will assume a conservative
limit of 10 contributing
resonances in the effective energy window for charged and neutral 
particle-induced reactions.

Fixing the required number of levels within the energy window of width $\Delta$,
one can find the minimum
temperature at which the above described condition is fulfilled.
Those temperatures (above which the statistical model can be used)
are plotted in a logarithmic color scale in Figs.~\ref{neutcrit}--\ref{alph}.
For neutron-induced reactions Fig.~\ref{neutcrit} applies,
Fig.~\ref{prot} describes proton-induced reactions, and
Fig.~\ref{alph} alpha-induced reactions.
Plotted is always the minimum stellar temperature $T_9$ (in 10$^9$ K) for the 
compound nucleus of the reaction.
It should be noted that the derived temperatures will not change considerably
even when changing the required level number within a factor of about two,
because of the exponential dependence of the level density on the
excitation energy.

This permits to read directly from the
plot whether the statistical model cross section can be ``trusted'' or
whether single resonances or other processes (e.g.\ direct reactions) 
have also to be
considered. (However, this does not necessarily mean that the
statistical cross section is always negligible in the latter cases,
since the assumed condition is quite conservative).
The above plots can give hints on when it is safe to use the statistical 
model approach and which nuclei have to be treated with special 
attention for specific temperatures. Thus, information on which nuclei 
might be of special interest for an experimental investigation may also 
be extracted.

\section{Summary}

\noindent
In the first part of the paper we described the most recent approaches 
being used for the application of statistical model calculations in 
astrophysical applications. In the second part we focussed 
on the level density description which contained the 
largest error when using the properties 
described before.
We were able to improve considerably the prediction of
nuclear level densities by employing an energy-dependent description for the
level density parameter $a$ and by properly including microscopic
corrections and back-shifts.
All nuclei can now be described with a single
parameter set consisting of just three parameters. 
The globally averaged deviation of prediction from experiment of about 
1.5 translates into a somewhat lower error in the final cross 
sections due to the dominance of transitions to states with low 
excitation energies. 
This will also
make it worthwile to recalculate the cross sections and thermonuclear
rates for many astrophysically important reactions in the intermediate 
and heavy mass region.

Finally, we also presented a ``map'' as a guide for the application of 
the statistical model for neutron-, proton- and alpha-induced 
reactions. Figs.~\ref{neutcrit}, \ref{prot}, \ref{alph} (as well as 
Figs.~\ref{hcomp}, \ref{etfsicomp}) as full size color plots can be obtained
from the first author.
The above plots can give hints on when it is safe to use the statistical
model approach and which nuclei have to be treated with special
attention at a given temperature. Thus, information on which nuclei
might be of special interest for an experimental investigation may also
be extracted.
It should be noted that we used very conservative assumptions in deriving
the above criteria for the applicability of the statistical model.

\acknowledgements

We thank F. K\"appeler and co-workers for making a preliminary version of
their updated neutron-capture cross section compilation available to us.
We also thank P. M\"oller for discussions.
This work was supported in part by the Swiss Nationalfonds.
TR is acknowledging support by the Austrian Academy of Sciences.

%


\begin{figure}
\caption{\label{figrat}Ratio of predicted to 
experimental~\protect{\cite{ilji92}} level 
densities at the neutron separation energy. The deviation is less than a 
factor of 2 (dotted lines) for the majority of the considered nuclei.}
\end{figure}

\begin{figure}
\caption{\label{hilffit}Ratio of predicted to 
experimental~\protect{\cite{ilji92}} level
densities at the neutron separation energy when using microscopic corrections
from the Hilf mass formula~\protect{\cite{Hilf76}}.}
\end{figure}

\begin{figure}
\caption{\label{hcomp}Ratio of the level density at the 
neutron separation
energy calculated with microscopic corrections from 
Hilf et al~\protect{\cite{Hilf76}} to those
calculated using corrections from FRDM~\protect{\cite{moell95}}.}
\end{figure}

\begin{figure}
\caption{\label{etfsicomp}Ratio of the level density at the 
neutron separation
energy calculated with microscopic corrections from
ETFSI~\protect{\cite{pearson}} to those
calculated using corrections from FRDM~\protect{\cite{moell95}}.}
\end{figure}

\begin{figure}
\caption{\label{bakaplot}Ratio of theoretical to 
experimental~\protect{\cite{bakanew}}
neutron capture cross sections at 30 keV. Cross sections for light nuclei 
($A<30$) are
not plotted because the statistical model cannot be applied in that region
for neutron-capture reactions (compare Fig.~\protect{\ref{neutcrit}}).}
\end{figure}

\begin{figure}
\caption{\label{levcrit}Deviation of the integration over the level density
from the exact sum over levels, depending on the number of levels in the 
energy window. The full line describes a ``worst case'' with narrow
resonances, the dotted line applies for broader resonances (e.g neutron
s-waves).}
\end{figure}

\begin{figure}
\caption{\label{neutcrit} (Color) Stellar temperatures (in 10$^9$ K) for 
which the statistical
model can be used. Plotted is the compound nucleus of the neutron-induced
reaction n+Target. Stable nuclei are marked.}
\end{figure}

\begin{figure}
\caption{\label{alph} (Color) Stellar temperatures (in 10$^9$) for which the 
statistical model can be
used. Plotted is the compound nucleus of the alpha-induced reaction
alpha+Target. Stable nuclei are marked.}
\end{figure}

\begin{figure}
\caption{\label{prot} Stellar temperatures (in 10$^9$) for which 
the statistical 
model can be used. Plotted is the compound nucleus of the proton-induced 
reaction p+Target. Stable nuclei are marked.}
\end{figure}

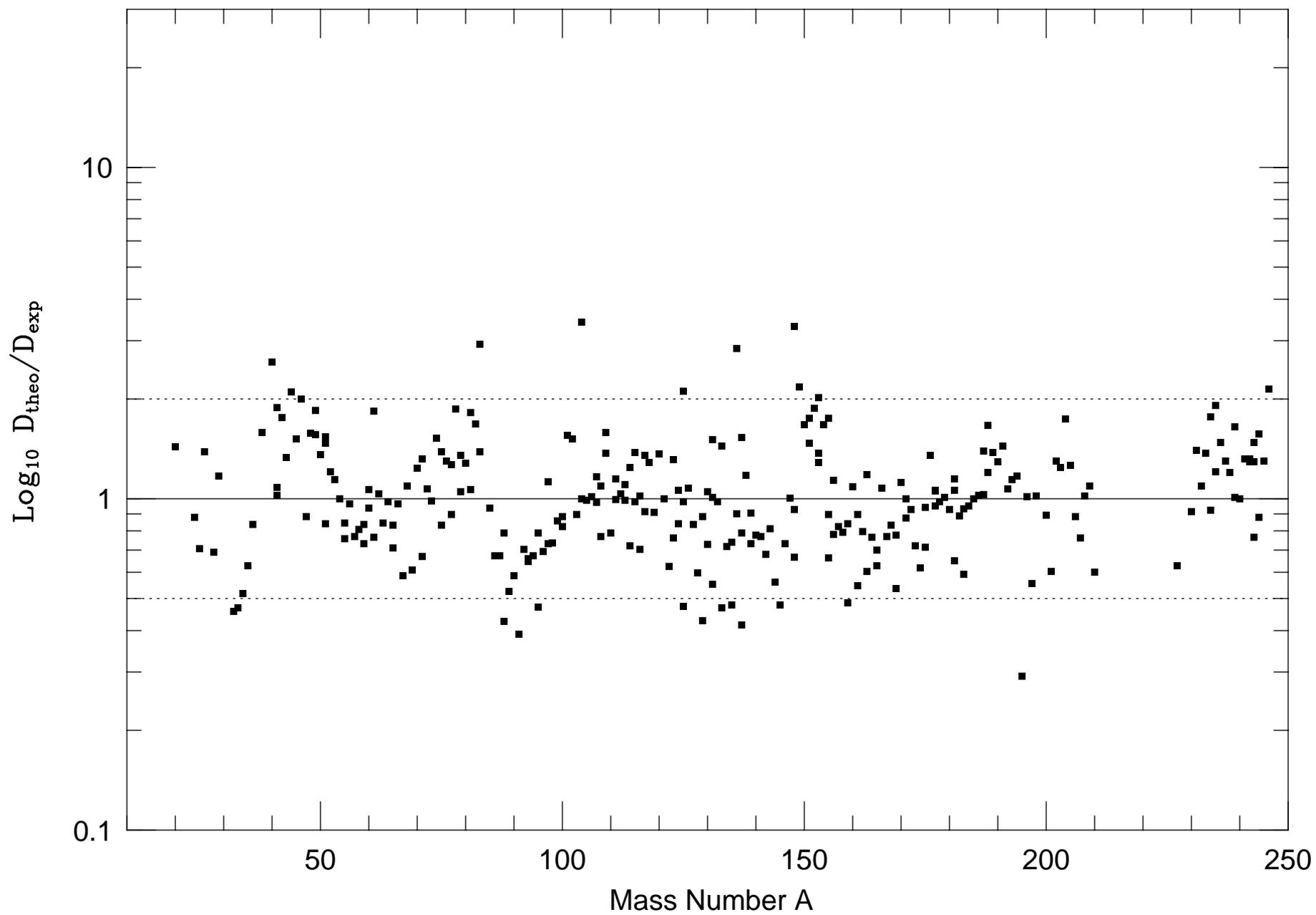
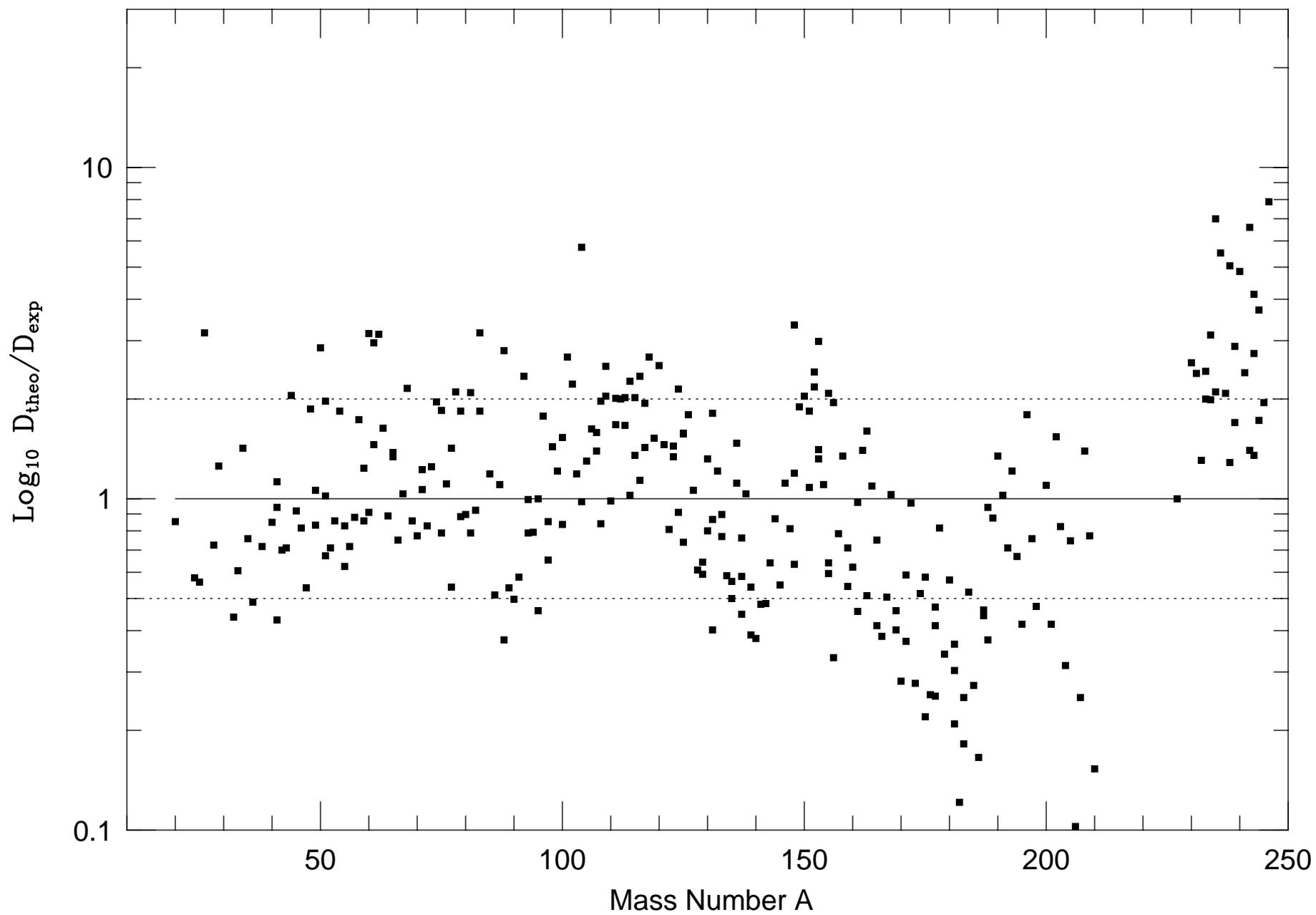
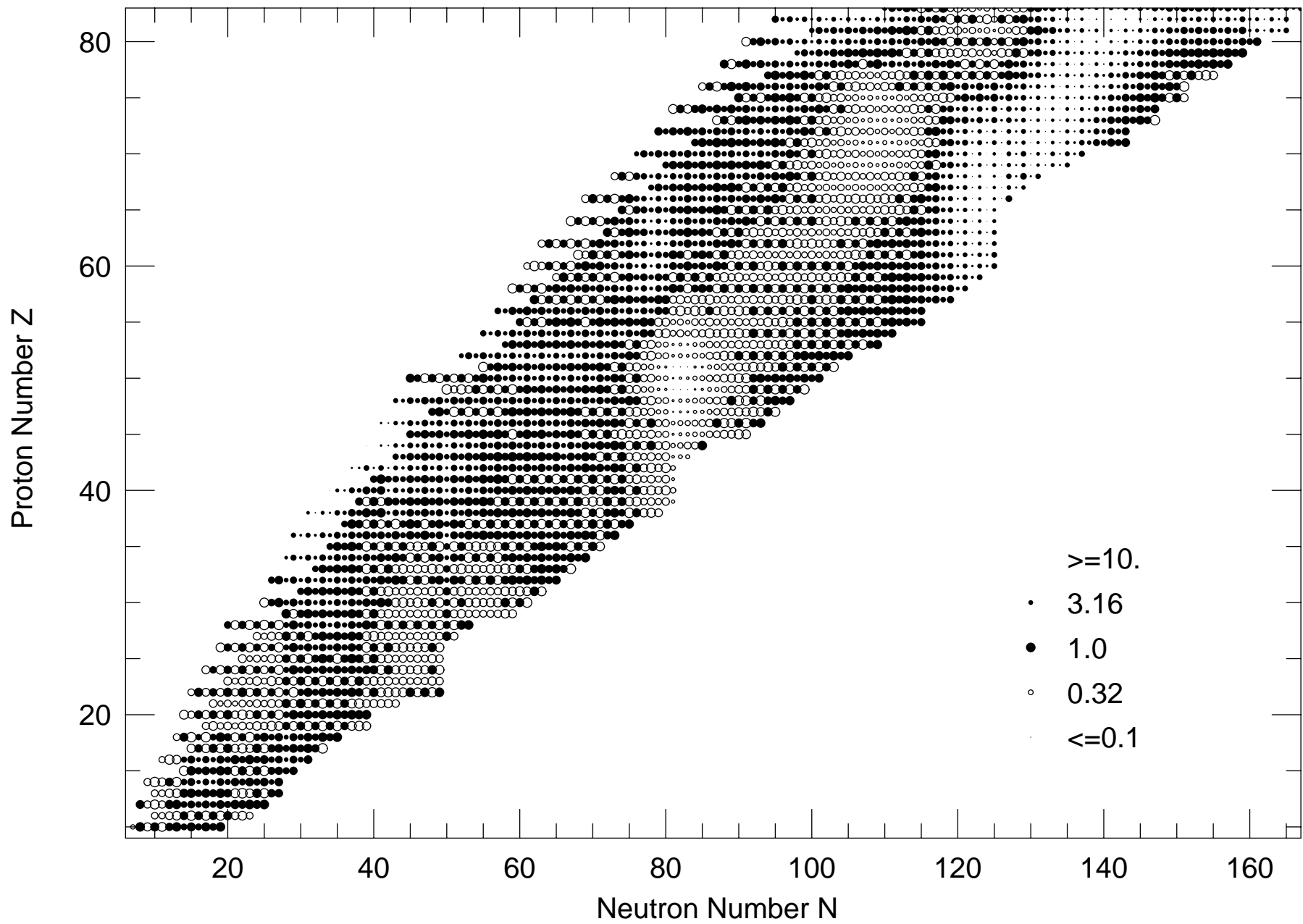
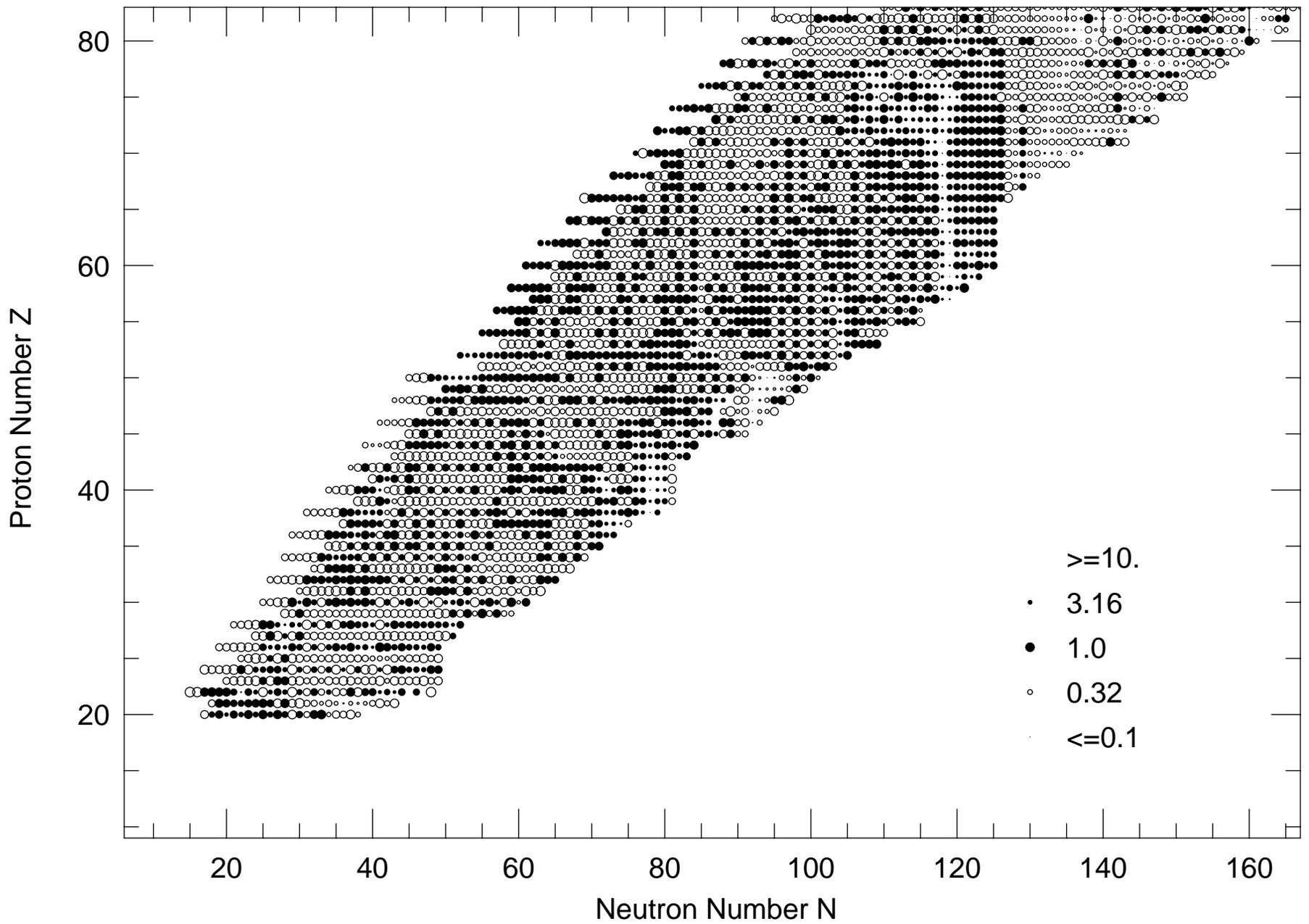
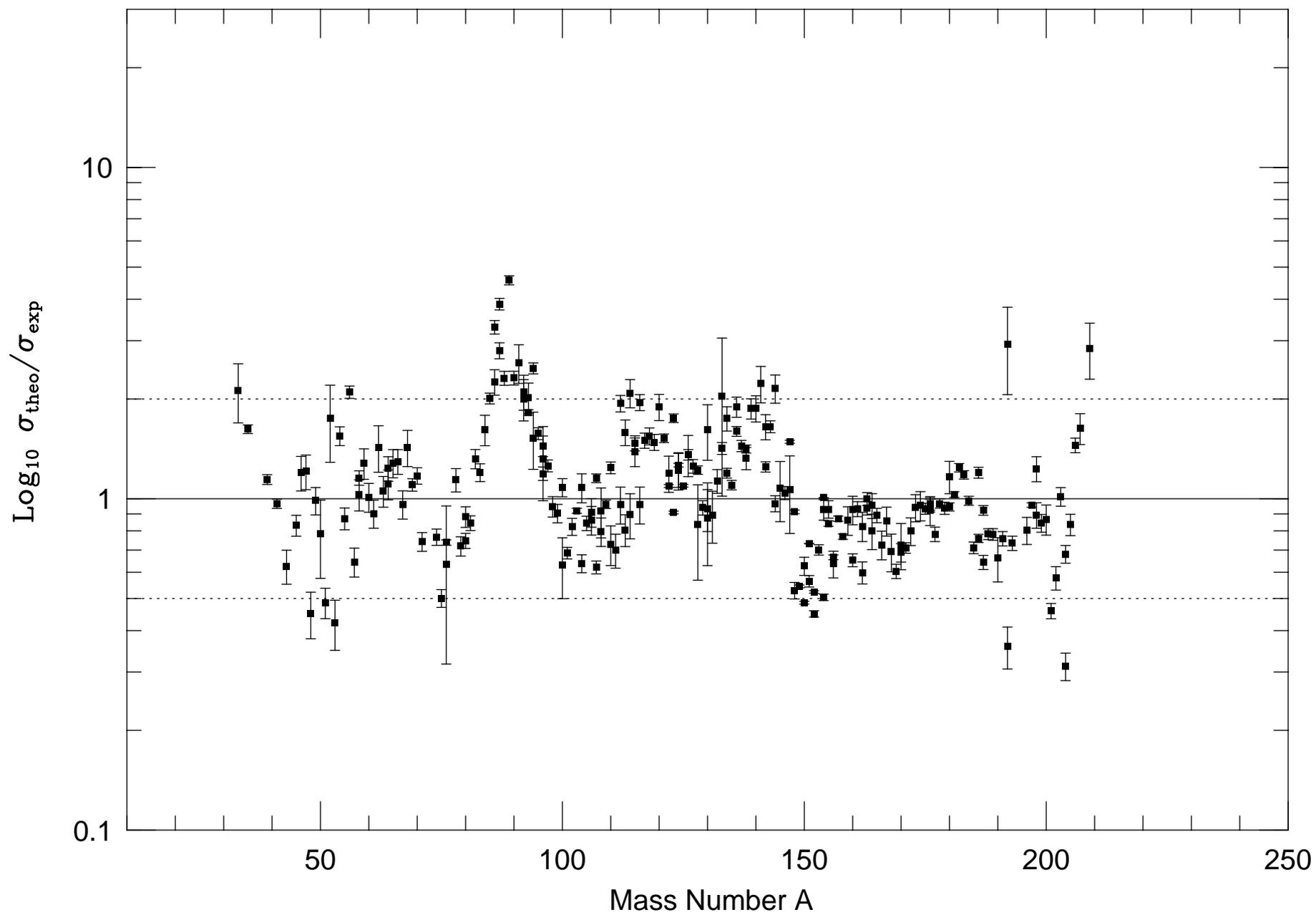
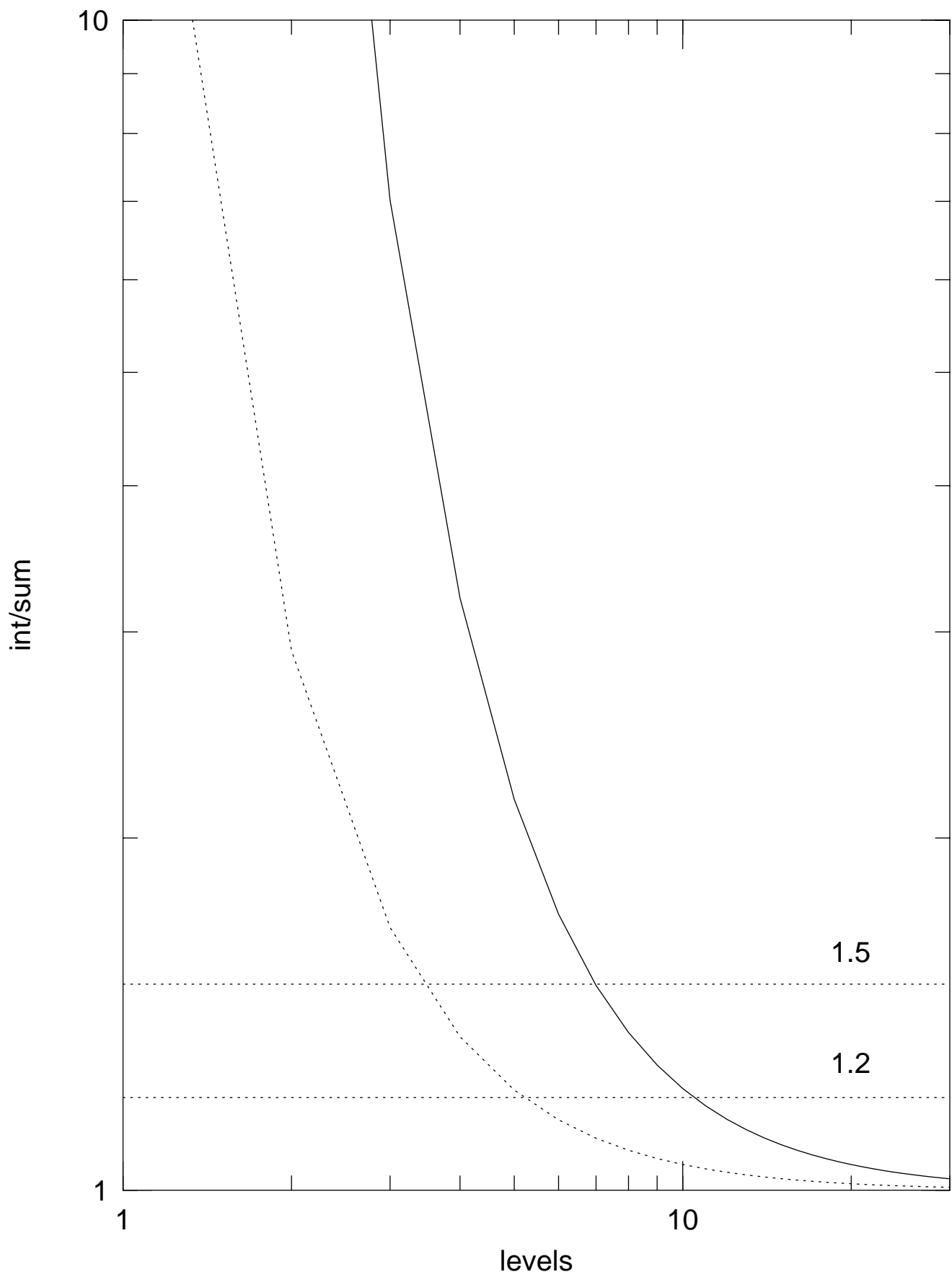
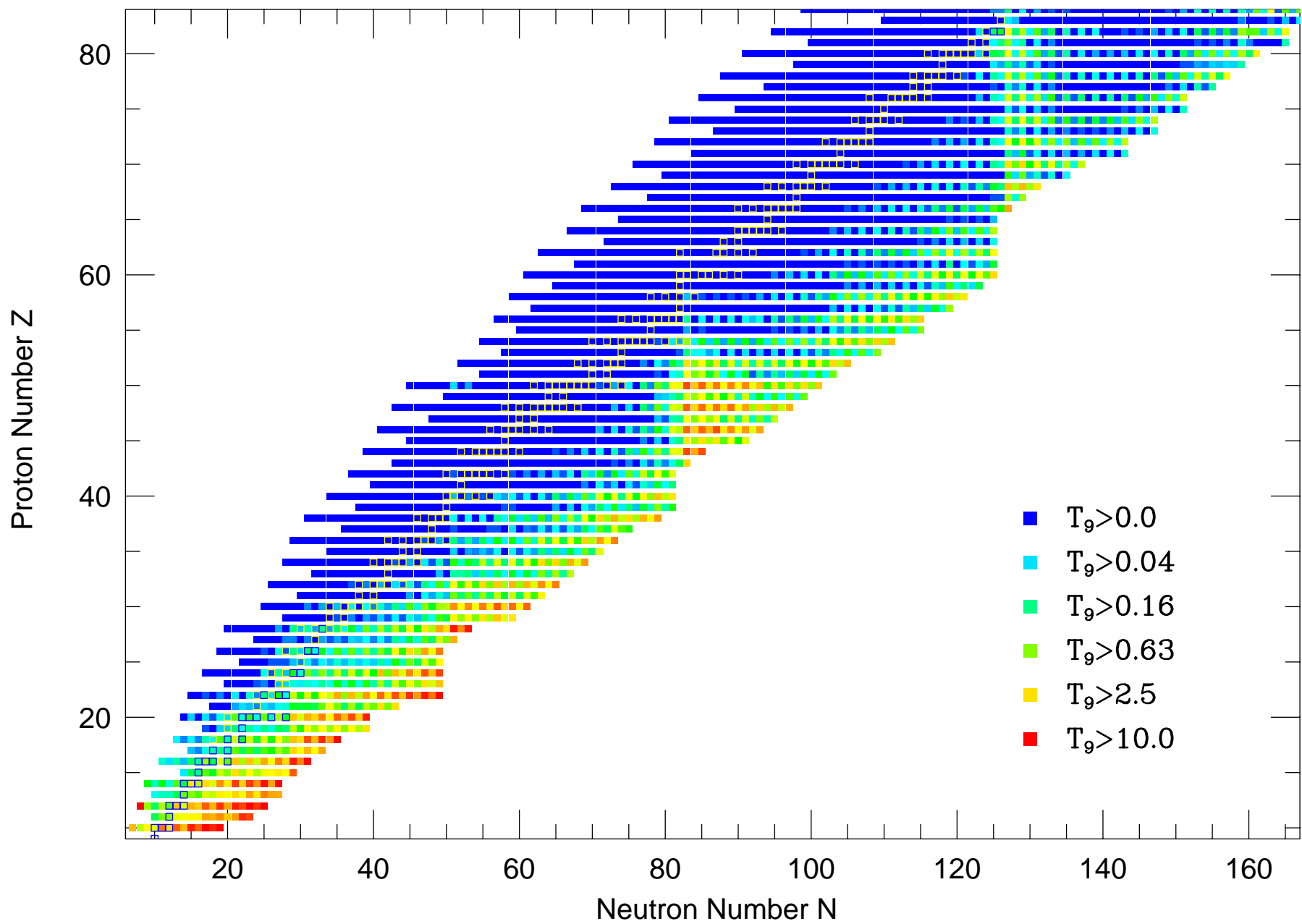
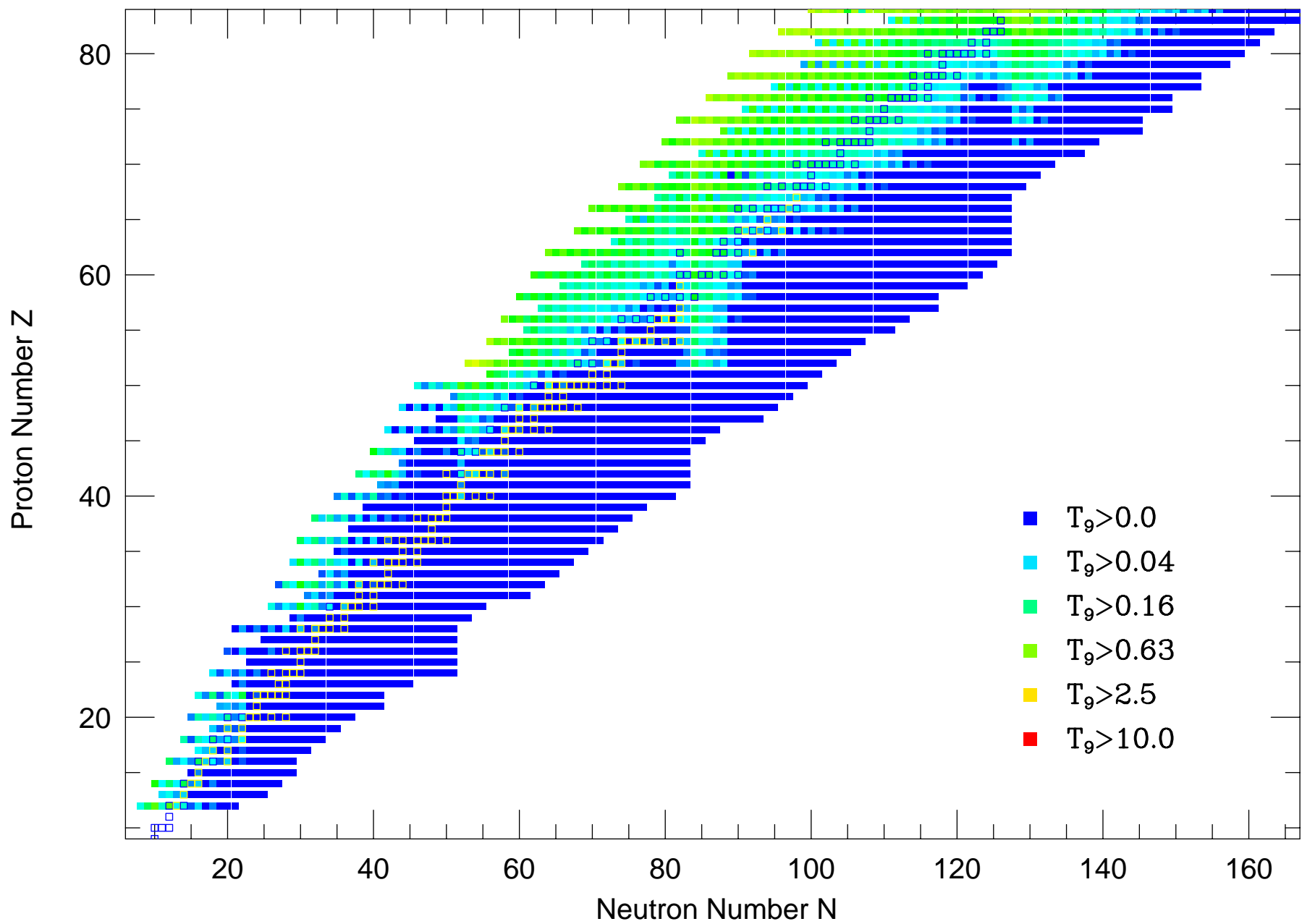
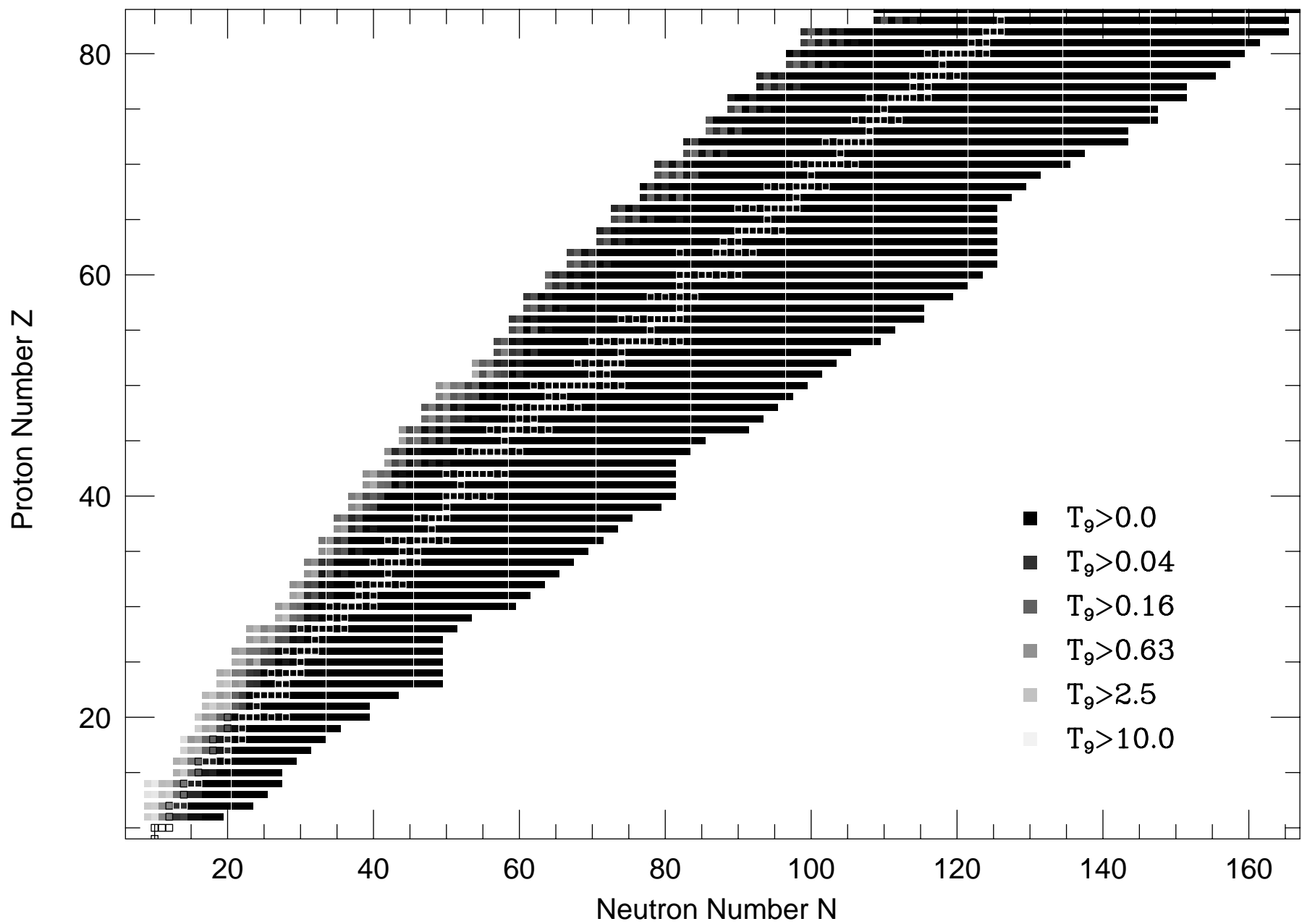
\end{document}